# Single-molecule observation of DNA replication repair pathways in *E. coli*

**Adam J. M. Wollman, Aisha H. Syeda, Peter McGlynn, Mark C. Leake**

**Abstract**

The method of action of many antibiotics is to interfere with DNA replication – quinolones trap DNA gyrase and topoisomerase proteins onto DNA while metronidazole causes single and double stranded breaks in DNA. To understand how bacteria respond to these drugs, it is important to understand the repair processes utilised when DNA replication is blocked. We have used tandem *lac* operators inserted into the chromosome bound by fluorescently labelled *lac* repressors as a model protein block to replication in *E. coli*. We have used dual-colour, alternating-laser, single-molecule narrowfield microscopy to quantify the amount of operator at the block and simultaneously image fluorescently labelled DNA polymerase. We anticipate use of this system as a quantitative platform to study replication stalling and repair proteins.

1.Introduction

1.1 Antibiotics interfere with DNA replication

Different types of antibiotic kill bacteria by interfering with DNA replication. In bacteria a sophisticated complex of protein machinery, called the replisome, replicates DNA by unwrapping its double helix and using the two exposed single strands as templates for DNA synthesis, creating a structure called the replication fork.(Reyes-Lamothe et al. 2010) Failure to copy DNA completely or accurately results in potentially disastrous consequences for the cell. The antibiotic family of Quinolones bind to two bacterial complexes associated with DNA replication, DNA gyrase and topoisomerase IV.(Mustaev et al. 2014) These complexes remove positive DNA supercoils generated by the replisome and also disentangle intertwined sister chromosomes as replication proceeds. These reactions occur by binding of gyrase or topoisomerase IV to one DNA duplex, cleavage of that duplex and passage of another region of the chromosome through the break prior to sealing the break to reform an intact chromosome. The outcome of this complex reaction is the release of torsional strain or chromosome disentanglement.(Drlica and Zhao 1997) Quinolones trap these topisomerases on the DNA by stabilising a covalent protein-DNA complex that is a normal part of the reaction cycle for both gyrase and topoisomerase IV, generating a protein block to replication and disrupting the DNA architecture.(Mustaev et al. 2014) Another antibiotic, Metronidazole, disrupts replication by inducing single strand and double strand breaks on

DNA in anaerobic pathogens.(Edwards 1977) Metronidazole is readily reduced, creating DNA damaging compounds in anaerobes but is easily re-oxidised in aerobes. There is evidence of increased DNA repair in *Helicobacter pylori* when exposed to Metronidazole.(Goodwin et al. 1998)

1.2 Replisome response to blocks

DNA damage occurs naturally in *E. coli*, due to reactive oxygen species, chemicals and radiation causing double and single stranded breaks on DNA. There are also natural protein blocks to replication. Transcription occurs concurrently with DNA replication and as RNA polymerases are an order of magnitude slower than replisomes in bacteria, collisions can occur. (McGlynn et al. 2012) RNA polymerases can also become stalled on the DNA by template damage, leading to the build-up of many polymerases.(McGlynn et al. 2012) Thus the replisome encounters many blocks to replication during the normal cell cycle and has been shown to pause frequently.(Gupta et al. 2013) Many of these stalled replisomes can continue if the block is removed which is advantageous as reloading the replisome can lead to genome rearrangements(Syeda et al. 2014). However replisomes lose functionality over time(Yeeles and Marians 2011) and so replisome reloading mechanisms are required for when replisome barriers are not cleared sufficiently rapidly prior to the blocked replisome losing activity.(Duch et al. 2013)

DNA replication is initiated from *oriC* in a sequence-specific manner on the genome. However, the replisome can stall anywhere and so different reloading and re-initiation mechanisms are required for stalled replication forks that are DNA structure- rather than DNA sequence-specific. These mechanisms are not fully understood. Two proteins, PriA and PriC, can both reload the replicative helicase DnaB back onto replication fork structures. DnaB plays a central role in the replisome, unwinding the two DNA template strands and also acting as an organising hub for the entire replisome complex. PriA and PriC recognise different forked DNA structures that together represent all possible types of fork structure on the chromosome.(Yeeles et al. 2013) *priA* and *priC* can be separately deleted from the genome but a knock-out mutant of both is not viable, thus these repair pathways are essential for cell survival. The *in vivo* dynamics of these proteins is unknown and there is evidence that DNA at forks needs processing by other proteins to allow repair or bypassing block.(Lecointe et al. 2007; Atkinson and McGlynn 2009) There are also an accessory helicase, Rep, which promotes the movement of replisomes through protein blockages on DNA.(Guy et al. 2009; Boubakri et al. 2010)

It is therefore important to study blocks to replication not only to understand the effect of antibiotics but also to understand how DNA replication is successfully completed in the face of the many natural blocks to replisomes inside cells. To study stalled replication, we have used a model protein block to replication by inserting tandem binding sites (34 copies of *lacO*) for the *lacI* transcription factor into the *E. coli* genome and over expressing the LacI protein. The Lac repressor-operator complex mimics naturally occurring protein-DNA complexes and inhibits fork movement with an affinity typically encountered during genome duplication. Since the majority of forks continue through a single complex unhindered, multiple complexes are required to give detectable inhibition of fork movement.(Payne et al. 2006; Guy et al. 2009) Studying replisomes stalled at these blocks is an ideal problem for single molecule microscopy, as it requires observation of individual replication machineries at blocks in the natural cell environment and also the associated repair proteins.

1.3 Single-molecule Fluorescence Microscopy

Fluorescent protein fusions can act as reporters to provide significant insight into a wide range of biological processes and molecular machines. They can be used to gain insight into stoichiometry and architecture as well as details of molecular mobility inside living, functional cells with their native physiological context intact.(Lenn et al. 2008; Plank et al. 2009; Chiu and Leake 2011; Robson et al. 2013; Llorente-Garcia et al. 2014; Bryan et al. 2014) These fusion proteins can be used in conjunction with single-molecule narrowfield microscopy, and its similar counterpart Slimfield microscopy, as a versatile tool to investigate a diverse range of protein dynamics in live cells to generate enormous insight into biological processes at the single-molecule level. It has been used in *E. coli* to investigate DNA replication by determining the stoichiometry of the components of the bacterial replisome(Reyes-Lamothe et al. 2010) and the proteins involved in the structural maintenance of chromosomes.(Badrinarayanan et al. 2012)

In narrowfield microscopy, the normal fluorescence excitation field is reduced to encompass only a single cell and produce a Gaussian excitation field (~20 $\mu m^2$) with 100–1000 times the laser excitation intensity of standard epifluorescence microscopy. Using such intense illumination causes fluorophores to emit many more photons, greatly increasing the signal to noise. This allows millisecond timescale imaging of individual fluorescently labelled proteins in their native cellular environment. This time scale is fast enough to observe the diffusional motion of proteins and the dynamics which may occur around the replication fork.

We have labelled the *lac* operator replication block with a fluorescent *lac* repressor-mCherry fusion protein together with the DnaQ replisome component fused to the monomeric green fluorescent protein (GFP) allowing simultaneous imaging of the replisome and block. Using a bespoke narrowfield microscope, we have observed complexes of these proteins in live cells (see schematic in figure 1). To reduce the impact of autofluorescence caused by the blue GFP-excitation light, we have used high speed alternating laser excitation (ALEX) to alternately excite each fluorophore at high speed. This enables the relatively dim mCherry protein to be observed without autofluorescence contamination and co-localised with GFP at high speed. Using custom software(Miller et al. 2015; Wollman et al. 2015a), we can quantify the number of fluorescently labelled proteins present in molecular complexes. Here, we demonstrate quantification of a replisome component and model protein replication block and show simultaneous imaging of both in the same live cell.

## 2. Methods

### 2.1 generating fluorescent strains

#### 2.1.1 Construction of chromosomal *dnaQ-mGFP* fusion

To create a *dnaQ-mGFP* C-terminal fusion, a PCR fragment containing mGFP and a downstream kanamycin resistance cassette amplified from pDHL580(Landgraf et al. 2012) using primers oAS77 and oAS78 was recombineered as described(Datsenko and Wanner 2000) immediately downstream of *dnaQ* into PM300, a derivative of MG1655. The recombinants were selected for kanamycin resistance and successful integration was confirmed by PCR and subsequent sequencing. The verified strain was called AS217.

#### 2.1.2 Construction of *lacI-mCherry* fusion

A synthetic *lacI-mCherry* C-terminal fusion from plasmid pAS13 (Eurofins MWG operon synthesis) was subcloned into pBAD24 between *Nco*I and *Xba*I sites that placed it under the control of the arabinose inducible promoter. The presence and orientation of the construct was confirmed by restriction digestion, sequencing and phenotypic testing in a reporter strain. The verified plasmid was called pAS17.

#### 2.1.3 Construction of strain with Lac repressor array, *dnaQ-mGFP* and *lacI-mCherry*

*dnaQ-mGFP* was moved from AS217 to a laboratory stock strain AS249 carrying *lacO*$_{34}$(Payne et al. 2006) by phage P1-mediated transduction. The transductants were selected for kanamycin resistance and presence of the *dnaQ-mGFP* allele was confirmed by PCR. The resulting *lacO*$_{34}$ *dnaQ-mGFP* strain (AS271) was transformed with pAS17 to create a dual labelled strain with an inducible roadblock to replication.

### 2.2 Growing strains and inducing the Lac repressor

Single colonies from transformation of AS271 with pAS17 were grown in 5 ml Luria-Bertani (LB) ampicillin and Isopropyl-β-D-thiogalactopyranoside (IPTG) in 15 ml culture tubes overnight. 1 ml of the overnight culture was washed twice with 1X 56 salts and inoculated into 10 ml 1X 56 salts together with ampicillin, glucose for growth and arabinose for Lac repressor induction and grown to an A$_{650}$ of 0.4-0.6 (mid log phase). Concentrations of ampicillin, glucose, arabinose and IPTG were 100 µg/ml, 0.1%, 0.02% and 1 mM respectively. Cells from 1 ml of culture were resuspended in 100 µl of fresh 1X 56 salts medium for visualization.

### 2.3 fluorescence microscopy

#### 2.3.1 The microscope

Our bespoke inverted fluorescence microscope was constructed from a Zeiss microscope body using a 100x TIRF 1.49 NA Olympus oil immersion objective lens and a *xyz* nano positioning stage (Nanodrive, Mad City Labs). Fluorescence excitation used 50mW Obis 488nm and 561nm lasers, modulated using TTL pulses sent from National Instruments digital modulation USB module. A dual pass GFP/mCherry dichroic with 25nm transmission windows centred on 525nm and 625nm was used underneath the objective lens turret. The beam was reduced 0.5x, to generate an excitation field of intensity ~6 Wcm$^{-2}$. The beam intensity profile was measured directly by raster scanning in the focal plane while imaging a sample of fluorescent beads. A high speed camera (iXon DV860-BI, Andor Technology, UK) was used to image at 5ms/frame with the magnification set at ~50 nm per pixel. Laser emission was modulated such that each laser was on for 5ms in alternating frames to give a 10ms sampling time with 5ms exposure time. The camera CCD was split between a GFP and mCherry channel using a bespoke colour splitter consisting of a dichroic centred at pass wavelength 560 nm and emission filters with 25 nm bandwidths centred at 525 nm and 594 nm. The microscope was controlled using our in-house bespoke LabVIEW (National Instruments) software.

### 2.3.2 Preparing samples and obtaining fluorescence data

*E. coli* cells were imaged on agarose pads suffused with media.(Reyes-Lamothe et al. 2010) In brief, gene frames (Life Technologies) were stuck to a glass microscope slide to form a well and 500 µl 56 salts media plus 1% agarose was pipetted into the well. The pad was left to dry at room temperature before 5 µl *E. coli* culture was pipetted in 6-10 droplets onto the pad. This was covered with a plasma-cleaned glass coverslip and imaged immediately. For each sample 10-30 cells were imaged in fluorescence and brightfield.

## 2.4 Analysing the data

Single fluorescent proteins or complexes of proteins can be considered point sources of light and so appear as spatially extended spots in a fluorescence image due to diffraction by the microscope optics.(Wollman et al. 2015b) Narrowfield fluorescence microscopy data consists of a time-series of images of spots which require *in silico* analysis to track each spot. We used custom Matlab™ tracking software to automatically identify spots, quantify them and link them into trajectories.(Miller et al. 2015; Wollman et al. 2015a) The software identifies candidate bright spots using a combination of pixel intensity thresholding and image transformation. The threshold is set using the pixel intensity histogram as the full width half maximum of the peak in the histogram which corresponds to background pixels. A series of morphological transformations including erosion and dilation is applied to the thresholded image to remove individual bright pixels due to noise and leave a single pixel at each candidate spot co-ordinate. The intensity centroid of candidate spots is found using iterative Gaussian masking(Thompson et al. 2002) and the characteristic intensity is defined as the sum of the pixel intensities inside a 5 pixel radius region of interest around the spot minus the local background(Xue and Leake 2009) and corrected for non-uniformity in the excitation field. If this spot is above a pre-set signal to noise ratio – defined as the characteristic intensity divided by the standard deviation of the local background, it is accepted. Trajectories are formed by linking together spots in adjacent frames based on their proximity and intensity.

The number of fluorophores present in a molecular complex is determined by dividing its intensity by the intensity of a single fluorophore. The characteristic intensity of a single fluorophore can either be determined from *in vitro* measurements of purified fluorophore or from the *in vivo* data itself using the intensity of spots found after bleaching the cell.

3. Results and discussion

3.1 Quantifying DnaQ

The single labelled DnaQ-GFP strain was imaged using narrowfield microscopy. An example cell is shown in Figure 2. The brightfield image of the cell is shown in fig. 2a and the fluorescence image of DnaQ-GFP shown in fig. 2b. Two spots of DnaQ can be seen in the fluorescence image corresponding to the two copies of the replisome, consistent with previous observations.(Reyes-Lamothe et al. 2010) Spots found by software over all frames are shown as green and blue circles in fig. 2a with their intensity values plotted against time in fig. 2c in units of characteristic GFP intensity. The spots have a stoichiometry of 3 DnaQ-GFP per replisome, consistent with previous observations.(Reyes-Lamothe et al. 2010)

The distribution of DnaQ replisome stoichiometries was obtained from a kernel density estimation and is shown in fig. 3. The stoichiometry peaks at 2 and ranges up to 6 DnaQ per replisome. This agrees well with previous observations of 2-3 per replication fork. These forks appear to be overlapping when replication is initiated from the origin leading to the observation of double stoichiometries. These results combined with recent measures of the total copy number of DnaQ(Wollman and Leake 2015) are in good agreement with a previous study which labelled DnaQ with the Ypet fluorophore. It has been suggested that the fluorophore used in a fusion protein can effect the stoichiometry of native complexes(Landgraf et al. 2012) but here we observe no difference between Ypet and monomeric GFP fusions.

3.2 Quantifying LacI

We then imaged the *lac* operator blocks. This required optimisation of the growth conditions and expression levels as the fluorescently labelled *lac* protein is not endogenously expressed in this strain. Cells were grown in minimal media so that growth is slowed and there is, on average, only one replisome per cell. This not only eliminates the noise caused by LB autofluorescence but also the signal from multiple replisomes. Thus having a single replisome greatly simplifies its tracking on the chromosome when it encounters the block and also makes downstream analysis easier by eliminating complexities due to multiple factors coming into play.

The results are shown in Figure 4, with a brightfield image in fig. 4a and mCherry fluorescence image in fig. 4b. Two mCherry spots are seen in the fluorescence image, consistent with the *lac* operator sites having been replicated. All spots found over time are marked as red circles in fig. 4a and their intensity plotted over time in fig 4c in units of mCherry intensities. The stoichiometry of the complexes is much lower than the 34 possible sites on the DNA and is closer to 5-10. This is unlikely to be caused by low expression levels as there is a significant diffuse background in the cell from unbound LacI-mCherry molecules. These results imply that the *lac* operators are not saturated with repressor. Further study is needed to understand the basis of this lack of saturation. The expression level could be varied and the number of potential binding sites on the DNA changed.

3.3 Dual colour experiments

The *lac* operator block plus LacI-mCherry has been incorporated into the DnaQ-GFP strain and preliminary data obtained. Figure 5 shows brightfield and fluorescence micrographs of the dual labelled strain. Our intention is to use this strain as a platform to study stalled replication by observing the behaviour of the replisome as it encounters different blocks with varying numbers of *lac* operators. This system could also be used to study repair proteins and could be combined with three colour microscopy, labelling the DnaQ with CyPet, the repair protein with Ypet and retaining the mCherry labelled *lac* operator array.

4. Summary

We have used tandem *lac* operators inserted into the chromosome bound by fluorescently labelled *lac* repressors as a model protein block to replication in *E. coli*. This block is a model for the action of some antibiotics such as quinolones which trap gyrases and topoisomerases on DNA. We have used dual-colour, alternating-laser, single-molecule narrowfield microscopy to quantify the amount of operator at the block and simultaneously image fluorescently labelled DNA polymerase. This quantitative platform for studying replication stalling will underpin future investigations.

**Figure legends**

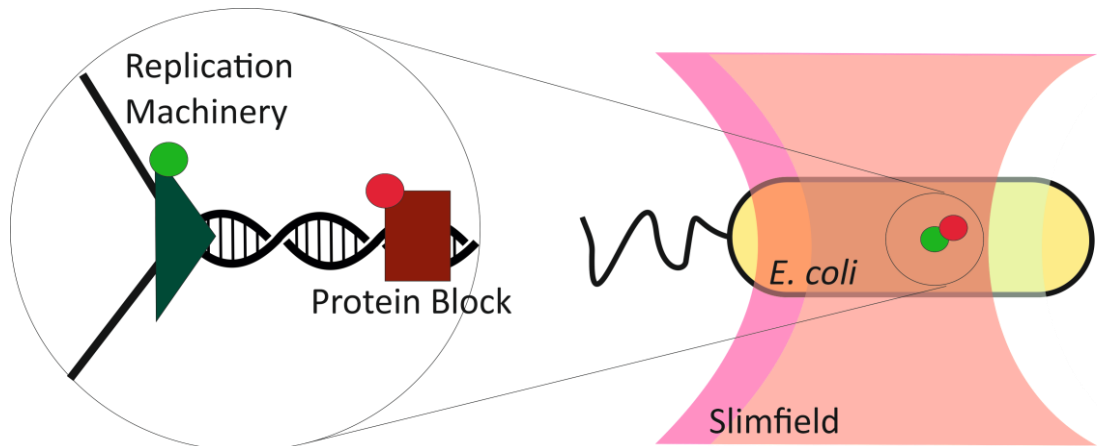

Figure 1: Schematic of slimfield observation of fluorescently labelled replisome components encountering a fluorescently labelled protein block in *E. Coli*.

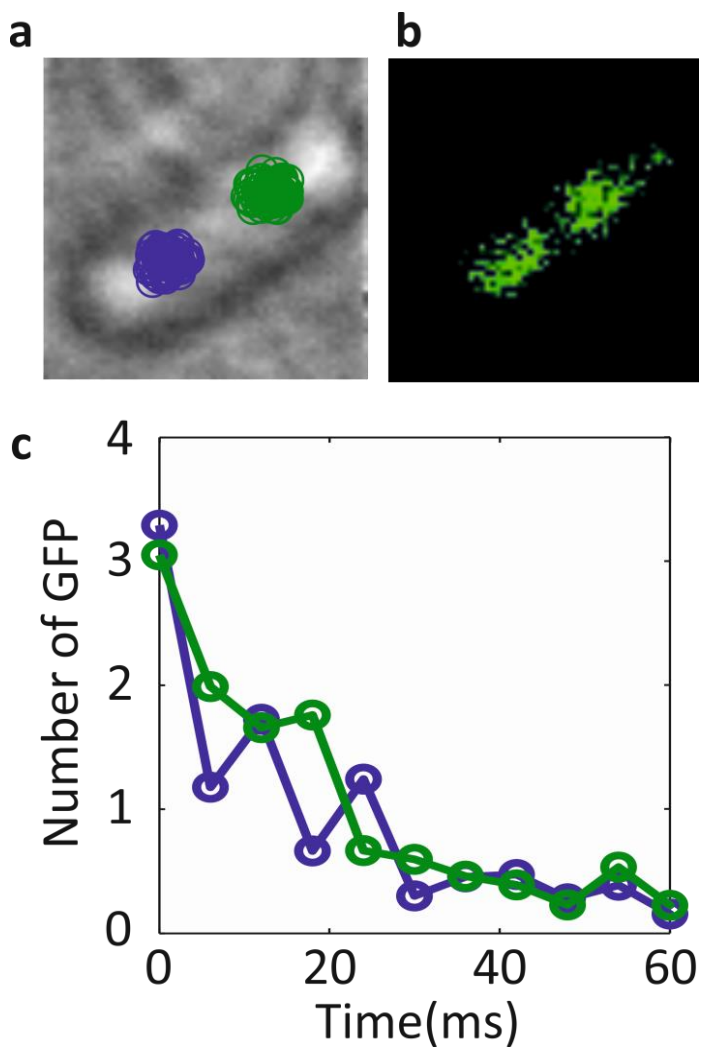

Figure 2: a – brightfield image of an *E. Coli* cell with tracked DnaQ-GFP overlaid, b – fluorescence micrograph of DnaQ-GFP, c – Intensity of each spot over time in GFPs

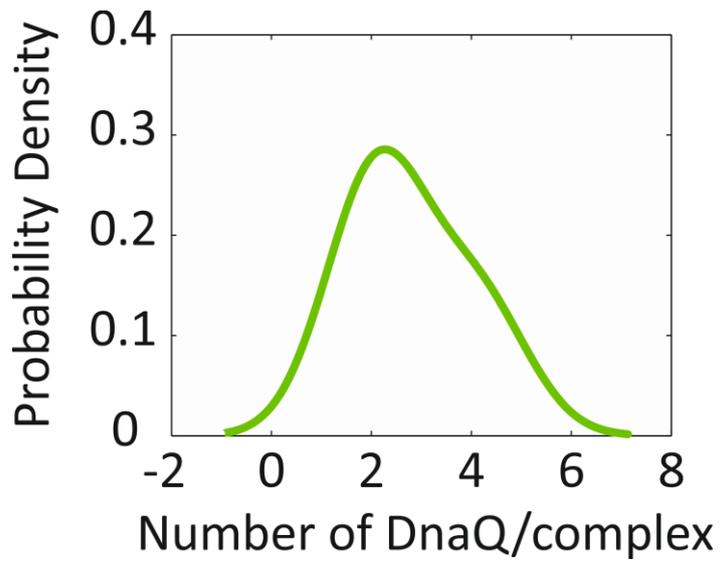

Figure 3: Kernal density estimation of the number of DnaQ-GFP per spot

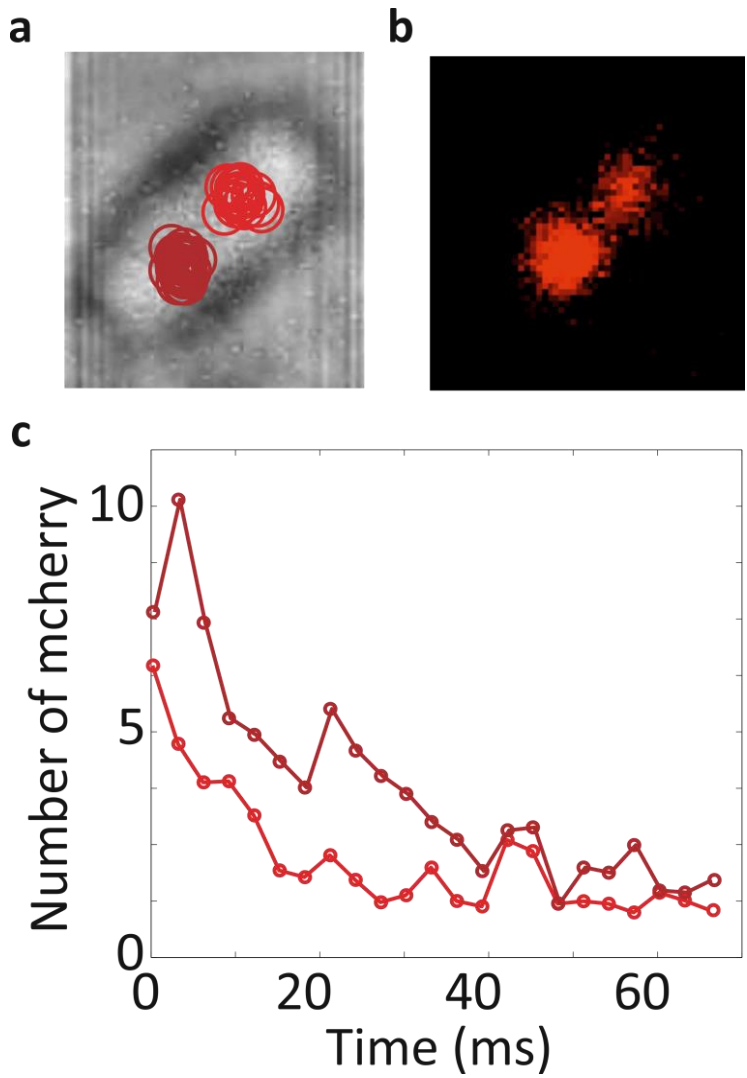

Figure 4: a – brightfield image of an *E. Coli* cell with tracked LacI-mcherry overlaid, b – fluorescence micrograph of LacI-mcherry, c – Intensity of each spot over time in mcherrys

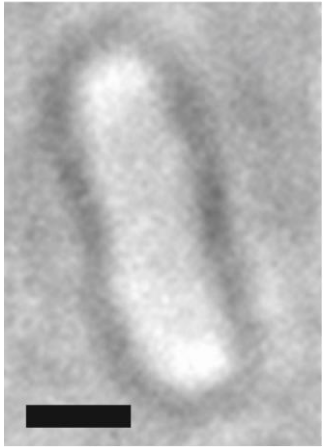 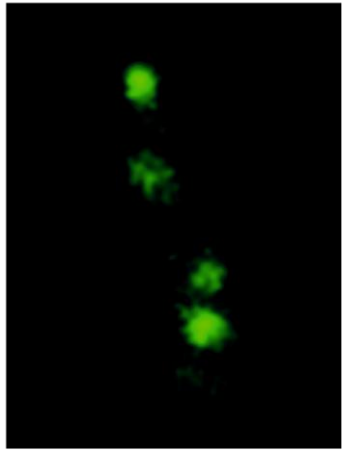 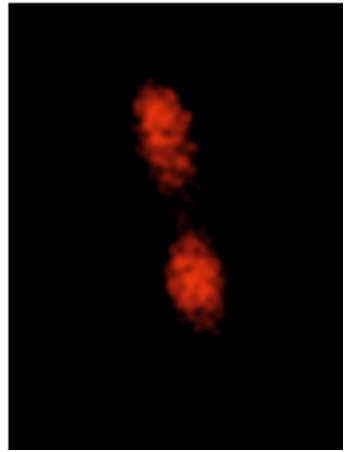

Figure 5: Left brightfield image of an E. Coli cell, middle fluorescence micrograph of DnaQ-GFP, right, fluorescence micrograph of LacI-mcherry in the same cell